\documentstyle[11pt]{article}

\def\afour{
\setlength{\topmargin}{0mm}
\setlength{\headheight}{0mm}
\setlength{\headsep}{0mm}
\setlength{\textwidth}{6in}
\setlength{\textheight}{248mm}
\setlength{\oddsidemargin}{.25in}
\setlength{\evensidemargin}{.25in}
}

\newcommand\eq[1]{Eq.~(\ref{#1})}

\newcommand\rfrac[2]{\left(\frac{#1}{#2}\right)}

\newcommand{\sub}[1]{_{\mbox{\scriptsize#1}}}

\newcommand\ee{\end{equation}}
\newcommand\be{\begin{equation}}
\newcommand\eea{\end{eqnarray}}
\newcommand\bea{\begin{eqnarray}}

\newcommand\sunit{\,\mbox{sec}}

\newcommand\km{\,\mbox{km}}

\newcommand\Mpc{\,\mbox{Mpc}}

\newcommand\mone{^{-1}}

\newcommand\half{^{1/2}}

\newcommand\msun{M_\odot}

\newcommand\del{{\mbox{\boldmath $\nabla$}}}
\newcommand\bfa{\mbox{\bf a}}

\newcommand\bfv{\mbox{\bf v}}

\newcommand\bfx{\mbox{\bf x}}

\newcommand\lsim{\mathrel{\rlap{\lower4pt\hbox{\hskip1pt$\sim$}}
    \raise1pt\hbox{$<$}}}
\newcommand\gsim{\mathrel{\rlap{\lower4pt\hbox{\hskip1pt$\sim$}}
    \raise1pt\hbox{$>$}}}

\def\calp{{\cal P}}

\afour

\begin{document}

\begin{flushright}
SUSSEX-AST 93/12-1 \\
LANCS-TH 9319\\
(December 1993, revised April 1994)\\
astro-ph/9401014\\
\end{flushright}
\begin{center}
\Large
{\bf Observational Constraints on the Spectral Index\\}

\vspace{.3in}
\normalsize
\large{David H. Lyth$^{\dagger}$ and Andrew R. Liddle$^*$} \\
\normalsize

\vspace{.6 cm}
{\em $^{\dagger}$School of Physics and Materials, \\ University of
Lancaster, \\ Lancaster LA1 4YB.~~~U.~K.}\\

\vspace{.4cm}
{\em  $^*$Astronomy Centre, \\ Division of Physics and Astronomy, \\
University of Sussex, \\ Brighton BN1 9QH.~~~U.~K.}\\

\vspace{.6cm}
Contribution to the 1993 Capri cmb workshop, to appear in\\
Astrophysical Letters and Communications

\vspace{.6 cm}
{\bf Abstract}
\end{center}

\vspace{.2cm}
\noindent
We address the possibility of bounding the
spectral index $n$ of primordial density fluctuations, using both
the cosmic microwave background (cmb) anisotropy, which probes scales $10^3$ to 
$10^4\Mpc$, and
data on galaxies and clusters which probes scales $1$ to $100\Mpc$.
Given $n$, sufficiently accurate large scale data on the cmb anisotropy 
can determine the normalisation of the primordial spectrum.
Then the small scale data are predicted within a given model of 
structure formation, which we here take as the MDM model determined
by the Hubble parameter $H_0$ and the neutrino fraction $\Omega_\nu$.
Each piece of small scale data is reduced to a value of $\sigma(R)$
(the linearly evolved {\em rms} density contrast with top hat smoothing 
on scale $R$) which allows data on different scales to be readily compared.
As a preliminary application, we normalise the
spectrum using the ten degree variance of the COBE data, and 
then compare the prediction with a limited sample of low energy
data, for various values of $n$, $\Omega_\nu$ and $H_0$.
With $H_0$ fixed at $50\km\sunit\mone 
\Mpc\mone$, the data constrain the spectral index to the range
$0.7\lsim n \lsim 1.2$.
If gravitational waves contribute to 
the cmb anisotropy with relative strength $R=6(1-n)$ 
(as in some models of inflation), the lower limit on $n$ is increased to
about $0.85$. The  uncertainty in $H_0$ widens this band
by about $0.1$ at either end.

\vfill
\begin{center}
E-mail addresses: lyth @ v1.phys.lancs.ac.uk ; 
arl @ starlink.sussex.ac.uk
\end{center}

\section*{Introduction}

A widely explored hypothesis about 
large scale structure is that it originates as an
adiabatic density perturbation, which is generated
during inflation as a vacuum fluctuation.
If this hypothesis is correct, the spectral index
of the perturbation provides a unique window on
the nature of the fundamental interactions at the inflationary energy 
scale, which is almost certainly many orders of magnitude higher than any 
scale that is directly accessible to either accelerator physics or astrophysics.
The reason is that it is determined by the shape
of the inflaton potential and the mechanism that ends inflation
(Davis et al. 1992; Liddle \& Lyth 
1992, 1993a, 1993b; Salopek 1992).
According to some models $n$ is a few percent less than 1, 
but there exist models where it is 
tens of percent below 1, others where it is
indistinguishable from 1, and at least one model where it 
can be ten to twenty percent {\it bigger} than 1 (Linde 1991;
Liddle \& Lyth 1993a; Copeland et al. 1994). 

The large angle cmb anisotropy provides a constraint on $n$ which is 
almost independent of any hypothesis about the dark matter or the value 
of the Hubble parameter (eg.~Wright et al., 1994; Gorski et al., 1994), but 
the constraint is weak because the data probe only the
decade of scales $10^3$ to $10^4\Mpc$.
We here address the possibility of a more accurate determination of
$n$ by combining the cmb anisotropy with 
data on galaxies and clusters, which probe the comparatively small 
 scales $1$ to $100\Mpc$.
The idea is to exploit the long lever arm provided by the simultaneous 
use of the two types of data.

Given $n$, sufficiently accurate large scale data on the cmb anisotropy 
can determine the normalisation of the primordial spectrum through the
Sachs-Wolfe effect.
Then the small scale data are predicted within a given model of 
structure formation, which we here take as the MDM model 
(Bonometto \& Valdarnini, 1984; Fang, Li \& Xiang, 1984; 
Shafi \& Stecker 1984; Klypin et al. 1993; 
Schaefer \& Shafi 1993), determined
by the Hubble parameter $h$ and the neutrino fraction 
$\Omega_\nu$.\footnote
{As usual $h$ is the Hubble 
constant in units $\Mpc\km\mone\sunit\mone$. 
The baryon density may be considered fixed through the nucleosynthesis 
prediction $\Omega_B h^2 = 0.013 \pm 0.002$ (Walker et al. 1991).}
Each piece of small scale data is reduced to a value of $\sigma(R)$
(the linearly evolved {\em rms} density contrast with top hat smoothing 
on scale $R$), allowing data on different scales to be readily compared.

As a preliminary application, we normalise the
spectrum using the ten degree variance of the COBE data, and 
present a limited set of low energy data which then constrain
$n$, $\Omega_\nu$ and $h$. 
We present results with 
$h=0.5$, and estimate roughly the effect of 
the uncertainty in $h$.
Finally we look at the effect of 
including a
gravitational wave contribution to the
cmb anisotropy, with relative strength $R=6(1-n)$ as predicted in
in some models of inflation. A more detailed study including
a full investigation of the 
effect of varying $h$  is in progress (Liddle et al. 1994).

Though differing in significant respects, the present work 
is not unrelated to early studies of the MDM model.
Pogosyan and Starobinsky (1993) have looked at the effect 
of varying $H_0$ and $\Omega_\nu$  with $n=1$. In addition, 
Liddle and Lyth (1993b) and Schaefer and  Shafi (1993) have 
varied $n$ and $\Omega_\nu$  with $h=0.5$; a comparison with these latter
results is made in the Conclusion. 

\section*{The observational constraints}

Several different types of observation constrain the theory, on scales
ranging from about $1\Mpc$ to $10^4\Mpc$. As usually presented 
the data refer to different quantities, so that one cannot plot
them on a single graph to exhibit their scale dependence. To
avoid this problem, we here focus on a single quantity
$\sigma(R)$, defined as the linearly evolved {\em rms} of the
density contrast, after smoothing with a top hat filter of radius $R$.
As we shall discuss, practically all of the data 
can be presented in terms of this
quantity which makes it preferable to the more widely used power 
spectrum $\calp(k)$.

In this section we present various observational determinations of
$\sigma(R)$. They are compared in Table 1 and Figure 1 with a benchmark 
theoretical model, taken to be the pure CDM model 
($\Omega_\nu=0$), normalised
to the $10^0$ COBE data described below 
and with the canonical parameter choices
$n=1$, $h=0.5$ and $\Omega_B=0.05$. 
Then in the next section
we see what parameters are needed to actually fit the data.
Throughout we use the transfer functions of Schaefer and Shafi (1993).

Except for the pairwise galaxy velocity, we focus exclusively on
data in the linear regime, defined by $\sigma(R,z)\lsim 1$
where $z$ is the redshift. On a given scale, linear theory is valid as long 
as $\sigma(R,z)\lsim 1$, which means that it is valid up to the present
epoch on scales $R\gsim 10h\mone\Mpc$. On these scales, the data themselves 
can be taken to refer to the linear quantity. Smaller scales require
special treatment, as described in Section E below.

\subsubsection*{A: The large scale cmb anisotropy} 

To normalise the amplitude we use the
{\it rms} of the anisotropy 
observed by COBE, smeared on the ten degree scale which corresponds to a linear 
scale $10^3 $ to $10^4\Mpc$. Its observed value from two years of COBE 
data (Bennett et al. 1994) is $\Delta 
T/T=(1.1\pm 0.1) \times 10^{-5}$, which with a Gaussian window function and a 
flat spectrum would correspond to an expected quadrupole $Q_2=15 \mu K$. The 
formal error is in fact less than the cosmic variance (10\% with a weak 
dependence on $n$), which must be added in quadrature to yield an estimate of 
the underlying power spectrum amplitude which has an uncertainty of about 13\%.
Corrections 
for the non-Gaussian beam profile and incomplete sky coverage raise the central 
value to 
$Q_2=17.4 \mu K$ (Wright et al. 1994a), so we adopt as the equivalent ten degree 
anisotropy for a Gaussian window the figure 
\be
\Delta T/T=(1.3\pm.1) \times 10^{-5} \label{cobe}
\ee

We note here an important caveat that although the interpretation of the $10$ 
degree variance both observationally and theoretically is rather simple, 
alternative means of analysing the data (Wright et al. 1994b, Gorski et al. 
1994) have recently given higher values for the normalisation which. 
If a higher value is
confirmed it would have an important impact on the constraints that can be 
obtained, shifting the allowed range of $n$ to be somewhat lower.

\subsubsection*{B: The distribution of galaxies and galaxy clusters}

The number density contrast $\delta_N(\bfx)$ is known fairly well for
IRAS galaxies, optical galaxies, radio galaxies and Abell galaxy 
clusters, out to a distance of several hundred Mpc.
Given the biasing hypothesis that we discuss in a moment, one can
deduce the mass density contrast $\delta(\bfx)$, and hence the
dispersion $\sigma(R)$. Alternatively, one can deduce the correlation
function $\xi(R)$. From a number of possibilities, we have chosen to 
use a recent analysis by Peacock and Dodds (1993), which combines a variety of 
data sets.

The dispersion and the correlation function 
are related to the underlying
power spectrum $\calp(k)$ (per unit logarithmic interval of wavenumber
$k$) by 
\bea
\sigma^2(R) &=& \int^\infty_0 W^2(kR) \calp(k) dk/k \label{sigma} \\
\xi(R) &=& \int^\infty_0 W(kR) \calp(k) dk/k \label{xi}
\eea
where $W$ is the `top hat' window function, and $\xi(R)$ is taken
to be the volume averaged quantity (Peacock \& Dodds 1993).
As with $\sigma(R)$, we take $\xi(R)$ and $\calp(k)$ to denote the
present linearly evolved quantities.

The biasing hypothesis is  
that for each type of object 
\be
\delta_N(\bfx)\simeq b_N \delta(\bfx)
\ee
where $b_N$ is a scale independent bias parameter. 
If linear evolution is valid, the
spectrum
$\calp\sub{N}$ observed
in redshift space is then related to the linearly evolved
spectrum $\calp$ of the density 
contrast in real space by
\be
\calp\sub{N} = b_N^2 \left[ 1 + \frac{2}{3} \frac{1}{b_N} + \frac15 
\frac{1}{b_N^2}\right]\calp
\ee
Towards the lower end of the linear regime $R\gsim 10h\mone\Mpc$
there are significant corrections, both to this formula and to the
linear evolution of $\calp$. After estimating them, one can deduce 
the ratios $b_I:b_0:b_R:b_A$, and (because the corrections are 
non-linear) one can 
also determine the overall normalisation, specified say by $b_I$. 

Peacock and Dodds estimate $b_I:b_0:b_R:b_A
=1:1.3:1.9:4.5$ for the ratios, and $b_I=1.0\pm0.2$ for the normalisation.
Assuming the central value $b_I=1.0$ they give an estimate of the
present value of the linearly evolved spectrum $\calp(k)$
over the range $k/h=0.01$ to $0.45\Mpc\mone$. 
The estimate was obtained from measured values of $\sigma(R)$
or $\xi(R)$ using the prescriptions 
\bea
\sigma(R) &=& \calp\half(k_R) \label{sigpd} \\
\xi(R) &=& \calp\half(\sqrt 2 k_R) \label{xipd}
\eea
where 
\be
k_R=\left[\frac12\Gamma\left(\frac{m+3}{2}\right)\right]^{1/
(m+3)}\frac{\sqrt5}{R} \label{kr}
\ee
and $m\equiv (k/\calp)(d\calp/dk)$ is the effective spectral index
evaluated in the benchmark CDM model. 
These formulae are obtained by taking $m$ constant, and using the 
approximation 
\be
W(kR)=\exp(-k^2R^2/10)
\ee
which is exact for $kR\ll1$.

We have used these prescriptions to
convert the estimates of $\calp(k)$
into estimates of $\sigma(R)$. Note that since the original data 
consisted of measurements of $\sigma(R)$ and of $\xi(R)$ 
any error in the prescription will tend to cancel
(it would cancel exactly if all of the data consisted of $\sigma(R)$
as opposed to $\xi(R)$). The results are shown in Table 1 and 
Figure 1.\footnote
{For clarity only every other point is given, and a couple of points in 
the nonlinear regime are dropped. The points that have been dropped
give little additional information, and it is not clear to what extent
nearby points are statistically independent.}
For each point, the inside error bar shows the 
fractional uncertainty in $b_I\calp\half$, and the full error bar
combines this in quadrature with the estimated $20\%$ uncertainty 
in $b_I$. One has the freedom to move the {\em entire set} of points
up or down by the same amount (on our logarithmic scale) 
within the full error bars, or to move {\em each point separately}
within its own inside error bar. 

Although this prescription relating $\sigma(R)$ to $\calp(k)$ is 
adequate on scales $\gsim 10\Mpc$, it is too dependent on the
shape of the transfer function to be useful on smaller scales.
As we shall see, data on such scales directly constrain 
$\sigma(R)$, which is our main reason for focusing on that quantity
rather than on $\calp(k)$ or $\xi(R)$.

\subsubsection*{C: The peculiar velocity field}

Since the peculiar velocity field $\bfv(\bfx)$ is the gradient of a potential 
in linear theory, it can be constructed in principle from the radial 
component observed through the ratio of Doppler shift to distance (Bertschinger 
\& Dekel 1989). Then in principle one can deduce the density contrast from the 
equation $\bfv=t\bfa$, which is equivalent to
\be
\del.\bfv=-4\pi G t \rho\delta \label{7}
\ee
Unlike the density contrast, the peculiar velocity can reasonably be 
assumed to be the same as that  of the underlying matter at least on 
large scales, which in principle makes it a better probe than the
galaxy correlation and dispenses with the biasing hypothesis.

In practice one still needs the hypothesis at present in order to obtain
really powerful results, which are obtained by comparing the density field 
obtained via velocities with that obtained via galaxy surveys.
A recent study (Dekel et al. 1993) concludes that at 95\% confidence level 
$0.5 < b_I < 1.3$. This is consistent with the above estimate $b_I=1.0\pm 0.2$,
and suggests that the {\it upper} limit of that estimate cannot
be increased much. As a result the lower limits on $\sigma(R)$
provided by the galaxy correlation data should be rather reliable.

Although the peculiar velocity alone does not yet give very
powerful results, it 
is  not completely useless. The standard way of utilising 
it is to calculate the 
theoretical {\it rms} of $\bfv$ for a random location, after smoothing
over a sphere of radius $Rh\mone\Mpc$, and compare with what we observe in the 
sphere 
around us. (There are several variants of this procedure, such as averaging
the radial component over a sphere.) This method can be used on the scale 
$R\simeq 20h\mone$ to $60h\mone\Mpc$, and according to Schaefer and Shafi 
(1993) 
it gives the estimate shown in Figure 1 when compared with the benchmark value, 
the uncertainty being dominated by the cosmic variance. The conclusion, shared 
by many earlier studies, is that there is broad agreement with the galaxy 
correlation result, but that the uncertainty is much bigger.

Ultimately the aim will be to use \eq{7} directly. A preliminary study 
been done by Seljak and Bertschinger (1993) reports 
$\sigma(R)=1.3\pm0.3$ at $R=8h\mone\Mpc$. On this scale
$\sigma\sub{cdm}(R)=1.22$ which leads to
$\sigma(R)/\sigma\sub{cdm}(R)=1.06\pm0.24$. This is too high 
to be consistent with the other estimates (Figure 1) and 
we shall not consider it further.

\subsubsection*{D: The galaxy cluster number density}

The average number density $n(>M)$ of clusters with mass bigger
than $M\sim10^{15}\msun$ gives information on a scale of order 
$10h^{-1}$ Mpc. Within linear theory one can estimate $n(>M)$
by considering the density contrast $\delta_R(\bfx)$, smeared over a
sphere of radius $R$ which encloses mass $M$.
(For a review of this procedure see Liddle and Lyth 
(1993a).) The well known Press-Schechter estimate starts with the assumption
that the matter in regions of space where the {\em linearly evolved}
quantity $\delta(R,\bfx)$ exceeds 
some critical value $\delta_c\simeq 1.7$ is bound into objects with mass 
$>M$ (the value of $\delta_c$ is motivated by a spherical collapse model, 
which gives $\delta_c=1.68$).
The Gaussian distribution gives the
fraction of space occupied by such regions, and multiplying it by a more 
or less unmotivated factor 2 leads to the Press-Schechter estimate
for the mass fraction bound into objects with mass bigger than $M$,
\be
\Omega(>M)=\mbox{erfc}\rfrac{\delta_c}{\sqrt 2\sigma(R)}
\,.
\label{275}\ee
An alternative prescription (Bardeen et al. 1986)
is to identify $n(>M)$ with the number density of the peaks of 
$\delta_R(\bfx)$ whose height exceeds $\delta_c$, which gives a roughly similar 
result. Yet another method is to run an N-body simulation of the 
collapse, which again gives 
roughly similar results, and suggests (Lacey \& Cole 1994)
that the appropriate value for
$\delta_c$ is within $20\%$ or so of the theoretically motivated $1.7$.
The equivalent value  with Gaussian  smearing
at a fixed value of $M$ is $\delta_c\simeq1.3$, as one finds both by direct
calculation of $\sigma(R)$ with the two filters (Liddle \& Lyth 1993a),
and by N-body simulation (Lacey \& Cole 1994, Efstathiou \& Rees 1988).

A recent study of the number density of Abell clusters using
these methods (White, Efstathiou \& Frenk 1993) gives $\sigma(R)= 0.57 \pm 
0.05$, at $R=8/h\Mpc$, which is compared with the benchmark in Figure 1. 
This estimate is lower than that obtained from the galaxy correlation, but
compatible with it in view of the uncertainties.

A different quantity that can be observed is
$n(>v)$ where $v$ is the velocity
dispersion of the constituents (virial velocity). It can be
converted into $n(>M)$ using a spherical collapse model as reviewed for
example by Liddle and Lyth (1993a). Using the Press-Schechter estimate
and ignoring uncertainties due to the spherical collapse model, several
authors have estimated $\sigma(R)$ by this method, most recently
Carlberg et al. (1993) who find $\sigma(R)=0.75\pm0.15$ at $8 h^{-1}\Mpc$,
in agreement
with the galaxy correlation estimate mentioned earlier.

The estimate just mentioned is actually at redshift $z\simeq0.3$,
which was allowed for by taking into account the linear evolution 
$\sigma\propto (1+z)\mone$. In the future high redshift
estimates of $n(>M)$ and $n(>v)$ for clusters will be very informative, but
at present the uncertainties involved are too large to permit
very definite conclusions.

\subsubsection*{E: The density of high-redshift objects}

Going down in scale from clusters to galaxies, linear evolution 
is still valid at high redshift even though it fails before the present.
As a result one can use the Press-Schechter estimate \eq{275}
or N-body simulations with linear initial conditions
to estimate the mass fraction $\Omega(>M,z)$ and compare it with observation.

One approach is to use the observed number
density $n(>M,z)$ of quasars, together with reasonably astrophysics,
to establish a lower bound on $\Omega(>M,z)$.
One such estimate (Haehnelt 1993) is $\Omega(>10^{13}\msun,4.0)
>1\times 10^{-7}$, which using the
Press-Schechter formula gives $\sigma(R,4.0)>0.33(\delta_c/1.7)$
at $R=3.3h\mone\Mpc$. Taking $\delta_c=1.7$ and the linear 
evolution $\sigma\propto (1+z)\mone$ appropriate for pure CDM, this
gives the bound on $\sigma(R)/\sigma\sub{cdm}(R)$ 
plotted in Figure 1. Allowing some hot dark matter 
tightens this bound because it gives less growth at early 
epochs, but the effect is not very big for $\Omega_\nu\lsim 0.3$. 

Potentially more restrictive bounds 
(Subra\-manian \& Pad\-man\-abhan 1994; Mo \&
Mir\-alda-Es\-cude 1994; Kauffmann \& Charlot 1994)
are provided by damped Lyman alpha 
systems,  which at these redshifts seem to contain a mass fraction
comparable to that of present day galaxies. For instance, data
presented by Wolfe (1993) indicate that at $z=3$ the baryon
mass fraction is bigger than $0.0023$. Dividing this by the average
baryon fraction $\simeq 0.05$ for the universe, this translates to
$\Omega(>M,3.0)>0.046$. To calculate the corresponding bound on
$\sigma(R)/\sigma\sub{cdm}(R)$ one needs the mass $M$ of the systems,
which is not well known. Using
$M=3\times10^{11}\msun$ corresponding to $R=0.5h\mone\Mpc$, 
one finds the bound $\sigma(R)/\sigma\sub{cdm}(R)>.60
(\delta_c/1.7)$, which is shown 
in Figure 1 for $\delta_c=1.7$. It is not 
terribly sensitive to $M$ in the range $10^{10}$ to $10^{12}
\msun$. 

Although they are potentially of great significance, these
small scale constraints 
should be viewed with caution at present.
One problem is that they involve astrophysics as well as 
direct observation;  for example in the case 
of damped Lyman alpha systems our assumption of a universal baryon 
fraction is clearly questionable. In addition there is the uncertainty
attached to the use of the Press-Schechter formula with the canonical value 
$\delta_c=1.7$, and also a possible inadequacy on small scales
of our adopted transfer function.

\subsubsection*{F: Pairwise galaxy velocity dispersion}

The observations presented are the most 
useful ones pertaining to the linear regime.
Additional information can be obtained by going to the 
non-linear regime and comparing with numerical simulations. The most
important quantity to consider is probably the pairwise
galaxy velocity dispersion.
According to a study of pure CDM by
Gelb, Gradwohl and Frieman (1993),
the scale explored by this statistic is actually roughly the same scale
$8h\mone\Mpc$ that we discussed earlier.
(This is supposed to come about through non-linear effects, 
the typical galaxy separation being an order of magnitude less). 
Compared with the benchmark normalisation, the normalisation 
required by non-linear simulations of the pairwise galaxy velocity
{\it with pure cold dark matter} is the one shown
in Figure 1. It is seen to be compatible with
the normalisation required by the cluster number density, 
but probably too low to be compatible with the galaxy 
correlation results. However, simulations done with
MDM (Klypin et al. 1993) show that for fixed $\sigma(R)$ the pairwise
velocity is decreased relative to the pure CDM prediction,
perhaps  allowing compatibility with the galaxy correlation result.

\begin{table}
\begin{centering}
\begin{tabular}{|c|c|c|c|}
\hline\hline
   $Rh$  &   $\sigma\sub{cdm}(R)$  &   $\sigma(R)/\sigma\sub{cdm}(R)$ &
Origin \\
\hline
   0.5  &   5.73  &   $ >0.60(\delta_c/1.7)$  & Damped Lyman alpha 
systems \\
   3.25 &    3.82 &    $ >0.43(\delta_c/1.7)$  & Quasars \\
   8.0  &   1.23  &     $0.32 \pm 0.08$ & Galaxy pairwise velocity \\
   8.0 &  1.23  &  $0.46 \pm 0.04$ & Cluster abundance, $z=0$ \\
   8.0 &  1.23 &  $0.61 \pm 0.13$ &  Cluster abundance, $z=0.3$ \\
   9.1 & 1.1 & $0.61 \pm  0.04 $ & Galaxy \& cluster correlation \\
   14.2 &    0.70 & $0.68 \pm 0.04 $ & Galaxy \& cluster correlation \\
   22.3 & 0.40  & $0.84 \pm 0.04 $ & Galaxy \& cluster correlation \\
   35.1 & 0.22 &  $0.82 \pm 0.05 $ & Galaxy \& cluster correlation \\ 
   56.4 & 0.11 &  $0.86 \pm 0.11 $ & Galaxy \& cluster correlation \\
  40.0 &  0.15 & $1.0^{+0.15}_{-0.28}$ &  Bulk flow\\
\hline\hline
\end{tabular}
\caption{The data set, which is discussed in detail in the text.
The galaxy and cluster correlation points assume 
a bias parameter $b_I=1.0$.}
\end{centering}
\end{table}

\subsubsection*{Summary}

Our data set is summarised in 
Figure 1. Before comparing it with theory, one has to ask whether the 
different data points are compatible. 

The only definite discrepancy is the low value coming from the pairwise
galaxy velocity, but as already discussed it may be raised by 
introducing hot dark matter, and the value of $R$ at which it should be 
applied is also rather uncertain. 

Although not actually discrepant, the lower limit on $\sigma(R)/
\sigma\sub{cdm}(R)$ at $R\sim 1h\mone\Mpc$ coming
from damped Lyman alpha systems will be puzzling if it turns out to 
as high as the one shown in Figure 1. Such a high limit 
would seem to imply a minimum (or at least an extremely sharp flattening)
for $\sigma(R)/\sigma\sub{cdm}(R)$ 
somewhere in the range $1\lsim R\lsim 10h\mone\Mpc$, which neither 
MDM nor any similar fix of the CDM model can provide.

\section*{Constraining the spectral index}

So far we have compared the data only with the benchmark CDM model.
Now we ask what, if any, regime of parameter space provides a fit
to the data, and in particular what range of $n$ is allowed.
For the reasons stated we ignore the pairwise galaxy velocity 
point (the lowest point at $8h\mone\Mpc$), and for 
the moment also the Lyman alpha bound (extreme left hand
point). 

To precisely delineate the allowed regime,
we would need a precise prescription as to what constitutes a fit. 
In the two earlier 
investigations, different viewpoints were taken in this respect.
The first (Liddle \& Lyth 1993b) used a data set even more limited than 
the one that we have exhibited, and simply demanded that the curve
pass within the error bars of every point. Such an approach cannot
be used if there are incompatible error bars, and is 
dangerous if some error bars are only just compatible. The 
second (Schaefer \& Shafi 1993)
considered a relatively full data set, somewhat akin to the set 
presented in Figure 1 but with many more points coming from galaxy
correlations, calculated the weighted mean square difference
$\chi^2$ between theory and observations, and 
drew contours in the
$n$-$\Omega_\nu$ plane. 
Although potentially useful, this procedure too is
somewhat problematical given the
present state of the data. In particular, it is not clear that
values of $\sigma(R)$ or $\xi(R)$ deduced from the galaxy correlation
on nearby scales are statistically independent as the $\chi^2$ analysis
assumes, even in regard to that part of the error not arising
from the uncertainty in the bias parameter (shown as inside error bars
in our Figures). 

Thus it is difficult, at the present time, to decide how best 
to formalise the notion of an acceptable fit. Instead of making such a 
decision, we offer in Figures 1 and 2 some representative curves.
On the basis of these, we argue that $n$ must lie within the advertised
bands, if the curve is not to pass far outside the error bars of at 
least one piece of data.

Recall that we are normalising the curves using the
COBE $10^0$ variance,
which with $n=1$ is equivalent to an {\it rms} quadrupole 
$Q_2=(17.4\pm2.3)\mu K$. 
Consider first the lower bound $n\gsim 0.7$. Figure 1 shows the prediction
with $n=0.7$, $\Omega_\nu=0$ and the central value of the normalisation.
The prediction is clearly too low, and
adding hot dark matter ($\Omega_\nu>0$) obviously makes things worse.
In Figure 2 the effect of raising the normalisation by 1-$\sigma$
is shown. Now the value $n=0.7$ is seen to be marginally acceptable
(excluding the two points already mentioned), 
though a better fit would clearly ensue if one added
some hot dark matter and increased $n$. In this context, note
particularly that the galaxy correlation points can be varied
randomly only within their {\em inside} error bars, implying
a significant positive slope (this is the famous result that
the canonical CDM model has too much power on small scales).
It is clear that a value of $n$ significantly below 0.7 cannot
be accommodated.

Next consider the upper bound $n\lsim 1.2$. In Figure 2
the prediction with $n=1.2$ is plotted, with the normalisation
{\em reduced} by 1-$\sigma$. Even with this reduction,
the theoretical curve is significantly higher than the data,
for any reasonable value of $\Omega_\nu$.
(If one takes seriously the cluster abundance point
it is clear that even $n=1.2$ is completely ruled out.)

So far we have discounted the possibility of a significant gravitational 
wave contribution to the cmb anisotropy.
Such a contribution is predicted by some
models of inflation (Davis et al. 1992; Liddle 
\& Lyth 1992, 1993a, 1993b; Salopek 1992). Of the models 
which are well motivated from particle physics, those which give
a significant contribution also have $n<1$, and in them
the normalisation of the density 
perturbation is reduced by a factor $[1+6(1-n)]^{-1/2}$ 
when gravitational waves
are taken into account.
In Figure 2, the effect of including this factor is shown for
$n=0.85$ and $\Omega_\nu=0$, with the COBE normalisation raised by
1-$\sigma$. Its slope is clearly too small, but if the slope is 
increased by reducing $n$ or by making $\Omega_\nu>0$ the 
normalisation will be too low. Thus, $n$ cannot be significantly
less than $0.85$ with the gravitational waves.

These conclusions are for the case $h=0.5$, which is the central value
in the range $0.4 \lsim h \lsim 0.6$ allowed by
Hubble's law and the age of the universe.
Increasing
$h$ by $0.1$ is roughly equivalent to reducing $n$ by $0.1$ and 
{\em vice versa}
(see eg.~Liddle and Lyth, 1993), so the uncertainty in
$h$ widens the allowed band for $n$ by roughly $0.1$.

\section*{Conclusion}

Although it is hampered at the present time by somewhat inadequate data,
the simultaneous use of the large angle cmb anisotropy and of data on galaxies
and clusters is a potentially very powerful tool for constraining
the spectral index of the primordial adiabatic density perturbation.
 The model dependence introduced by considering the 
galaxy and cluster data is likely to be more than compensated by the
extra range of scales explored (about four decades as opposed to
one decade for the large scale cmb anisotropy alone).
Provided that the density perturbation is capable of accounting
for both types of data, it may be possible in the forseeable 
future to determine $n$ to an accuracy of a few percent, which
would allow a unique window on nature of the fundamental interactions
responsible for  inflation.

\section*{Acknowledgements}

ARL is supported by the Royal Society, and 
acknowledges the use of the STARLINK computer system at the University of
Sussex. We thank Pedro Viana for helpful discussions.


\section*{References}
\frenchspacing
\begin{description}

\item Bardeen, J. M., Bond, J. R., Kaiser, N. and Szalay, A. S., 1986, 
	Astrophys. J. 304, 15.
\item Bennett, C. L. et al., 1994, ``Cosmic Temperature Fluctuations from Two
	Years of COBE DMR Observations'', COBE preprint.
\item Bertschinger, E. and Dekel, A., 1989, Astrophys. J. Lett. 336, L5.
\item Bonometto, S. A. and  Valdarnini, R., 1984, Phys. Lett. 103A, 369 (1984).
\item Carlberg, R. G. et al., 1993, 
      ``Mapping Moderate Redshift Clusters'', preprint.
\item Copeland, E. J., Liddle, A. R., Lyth, D. H., Stewart, E. D. 
      and Wands, D., 1994, to appear, Phys. Rev. D, 15th June.
\item Davis, R. L., Hodges, H. M., Smoot, G. F., Steinhardt, P. J., 
	and Turner, M. S., 1992, Phys. Rev. Lett. 69, 1856.
\item Dekel, A., Bertschinger, E., Yahil, A., Strauss, M. A., Davis, M., 
	Huchra, J. P., 1993, Astrophys. J. 412, 1.
\item Efstathiou, G. and Rees, M. J., 1988, Mon. Not. Roy. astr. Soc. 230, 5p.
\item Fang, L. Z., Li, S. X. and Xiang, S. P., 1984, Astron. Astrophys. 140, 77.
\item Gelb, J. M., Gradwohl, B.-A. and Frieman, J. A., 1993, 
      Astrophys. J. 403, L5.
\item Gorski, K. et al., 1994, ``On Determining the Spectrum of Primordial
	Inhomogeneity from the COBE DMR Sky Maps: II. Results of Two Year
	Data Analysis'', COBE preprint.
\item Haehnelt, M. G., 1993, Mon. Not. Roy. astr. Soc. 265, 727.
\item Kauffmann, G and Charlot, S., 1994,
	``Constraint on models of galaxy formation from the evolution of damped 
	Lyman alpha absorption systems'', Berkeley preprint.
\item Klypin, A., Holtzman, J., Primack, J. R. and Reg\"{o}s, E., 
      1993, Astrophys. J. 416, 1.
\item Lacey, C and Cole, S., 1994, ``Merger Rates in Heirarchical Models of 
	Galaxy Formation. II. Comparison with N-Body Simulations'', Oxford
	preprint.
\item Liddle, A. R. and Lyth, D. H., 1992, Phys. Lett. B291, 391.
\item Liddle, A. R. and Lyth, D. H., 1993a, Phys. Rep. 231, 1.
\item Liddle, A. R. and Lyth, D. H., 1993b, Mon. Not. Roy. astr. Soc. 265,
        379.
\item Liddle, A. R., Lyth, D. H., Schaefer, R. K., Shafi, Q. and Viana, P., 
1994,
	in progress.
\item Linde, A. D., 1991, Phys. Lett. B259, 38.
\item Mo, H. J. and Miralda-Escude, J., 1994,
	``Damped lyman alpha systems and galaxy formation'', Princeton preprint.
\item Peacock J. A. and Dodds, S. J., 1993, to appear, Mon. Not. Roy. astr. Soc.
\item Pogosyan, D. Yu. and Starobinsky, A. A., 1993, Mon. Not. Roy. astr. Soc.
        265, 507.
\item Salopek, D. S., 1992, Phys. Rev. Lett. 69, 3602.
\item Schaefer, R. K. and Shafi, Q., 1993, ``A Simple Model of Large Scale 
        Structure Formation'', Bartol preprint.
\item Seljak, U. and Bertschinger, E., 1993, ``Amplitude of Primeval
        Fluctuations from Cosmological Mass Density Reconstructions'', MIT
        preprint.
\item Shafi, Q. and Stecker, F. W., 1984, Phys. Rev. Lett. 53, 1292.
\item Subramanian, K. and Padmanabhan, T., 1994, 
	``Constraints on the models for structure formation from the abundance 
	of damped lyman alpha systems'', IUCAA preprint 5/94.
\item Walker, T., Steigman, G., Schramm, D. N., Olive, K. A. and Kang, H.-S.,
        1991, Astrophys. J. 376, 51.
\item White, S. D. M., Efstathiou, G. and Frenk, C. S., 1993, Mon. Not. Roy.
        astr. Soc. 262, 1023.
\item Wolfe, A. M., 1993, Annals of the New York Academy of Sciences
	688, 836.
\item Wright, E. L. et al., 1992, Astrophys. J. 396, L13.
\item Wright, E. L. et al., 1994a, Astrophys. J. 420, 1.
\item Wright, E. L., Smoot, G. F., Bennett, C. L. and Lubin, P. M., 1994b, 
	``Angular Power Spectrum of the Microwave Power Spectrum seen by the
	COBE Differential Microwave Radiometer'', COBE preprint 94-02.
\end{description}

\section*{Figure Caption}

{\em Figure 1.}\\ 
Observation versus theory, described in detail in the text. The theoretical 
curves are all COBE normalised and given as the ratio with respect to the 
benchmark CDM model. The observational data points are
as follows. Lower limits: damped Lyman alpha systems (leftmost point) and 
quasars. Star: galaxy pairwise velocity dispersion. Open triangle: cluster 
number density (error bar angled only for visual clarity). Filled triangle: 
cluster velocity dispersion. Cross: bulk flow 
in spheres around us. Squares, galaxy and galaxy 
cluster correlation functions; 
the inner error bars on the squares are without the uncertainty 
in $b_I$ (see text for details). 

\vspace{20pt}
\noindent
{\em Figure 2.}\\ 
Similar to figure 1, showing some extreme choices of parameters, but with $h$ 
kept at $0.5$. These models are {\em not} COBE normalised; instead the COBE 
normalisation is allowed to shift by $1$-sigma (13\% 
with the incorporation of cosmic variance) in either direction to improve 
agreement with the data. Note that the introduction of hot dark matter is 
insufficient to compensate for tilt to $n = 1.2$, indicating a strong upper 
limit. For $n$ as low as $0.70$, adding hot dark matter only makes things 
worse. 
For models with significant gravitational waves, hot dark matter must be 
introduced to obtain the right shape for the galaxy correlation function, but 
the cost is excessively reduced short-scale power.

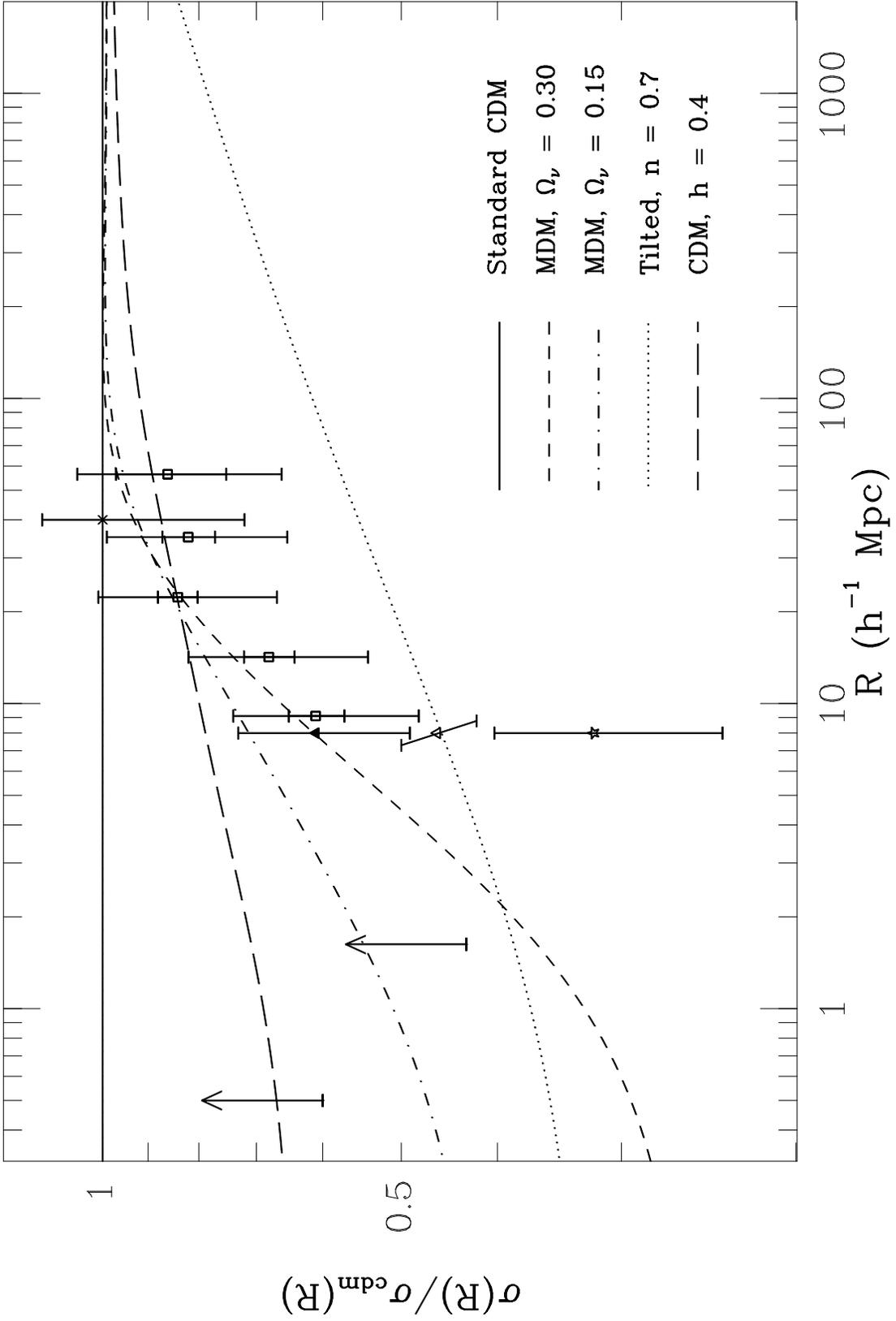
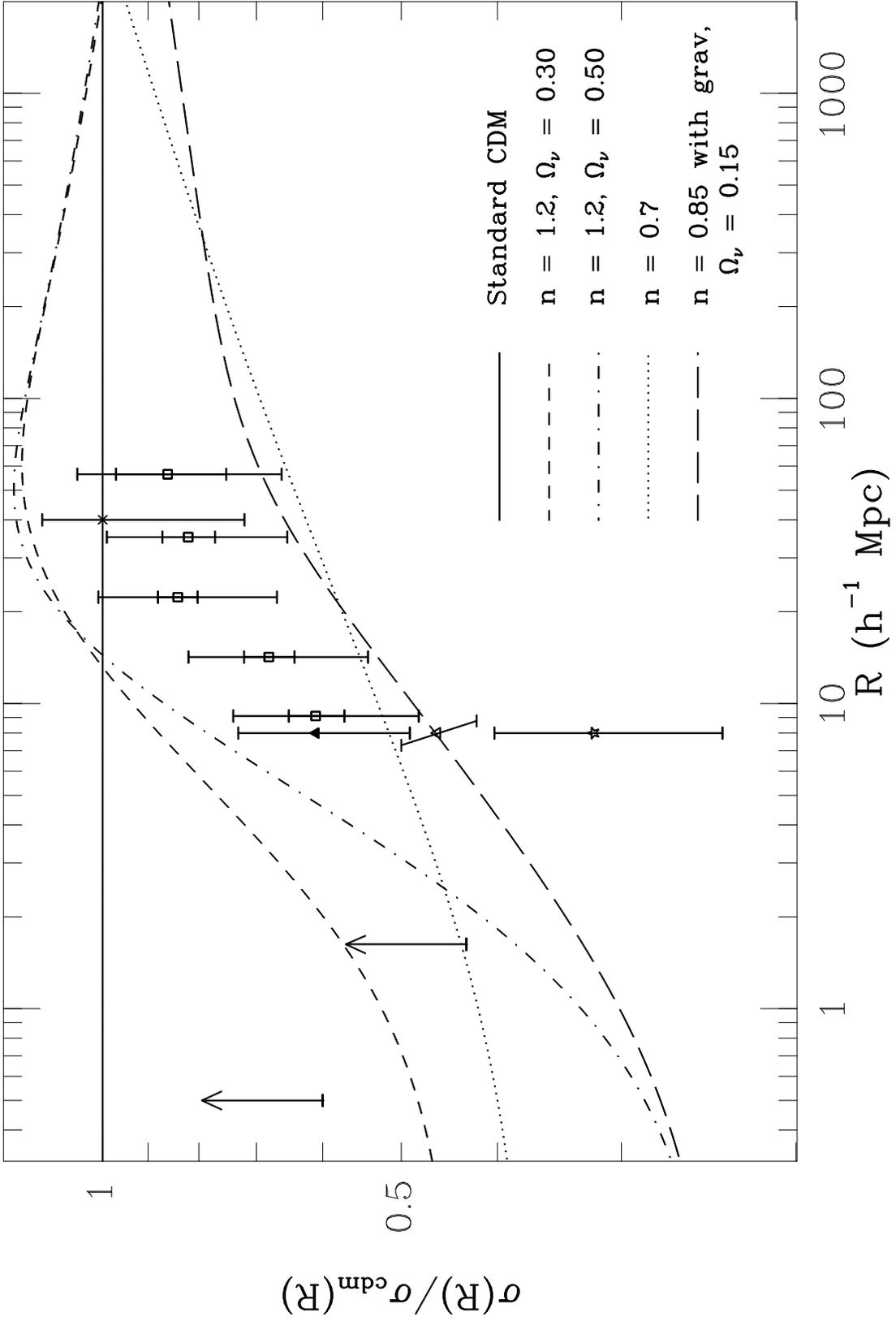
\end{document}